\documentclass {llncs}
\usepackage{rotating}
\usepackage{enumitem}
\usepackage{todonotes}
\usepackage{graphicx}
\usepackage{verbatim}
\usepackage{xcolor}
\usepackage[final]{pdfpages}

\newcommand{\specialcell}[2][c]{%
	\begin{tabular}[#1]{@{}l@{}}#2\end{tabular}}
\begin{document}

\title{A Conceptual UX-aware Model of Requirements}

\author{Pariya Kashfi\inst{1}, Robert Feldt\inst{1,2}, Agneta Nilsson\inst{1}, Richard Berntsson Svensson\inst{2}}

\institute{
	Software Engineering Division
	\\
	Department of Computer Science and Engineering
	\\
	Chalmers University of Technology and Gothenburg University
	\\
	\email{\{pariya.kashfi,robert.feldt,agneta.nilsson\}@chalmers.se} \and
	Software Engineering Research Lab\\
	School of Computing\\
	Blekinge Institute of Technology\\
	\email{\{robert.feldt, richard.berntsson.svensson\}@bth.se}
}

\maketitle
\begin{abstract}
	
User eXperience (UX) is becoming increasingly important for success of software products. 
Yet, many companies still face various challenges in their work with UX.
Part of these challenges relate to inadequate knowledge and awareness of UX and that current UX models are commonly not practical nor well integrated into existing Software Engineering (SE) models and concepts.
Therefore, we present a conceptual UX-aware model of requirements for software development practitioners.
This layered model shows the interrelation between UX and functional and quality requirements.
The model is developed based on current models of UX and software quality characteristics.
Through the model we highlight the main differences between various requirement types in particular essentially subjective and accidentally subjective quality requirements.
We also present the result of an initial validation of the model through interviews with 12 practitioners and researchers. 
Our results show that the model can raise practitioners' knowledge and awareness of UX in particular in relation to requirement and testing activities.
It can also facilitate UX-related communication among stakeholders with different backgrounds.

\end{abstract}							

\section{Introduction}
\label{sec:intro}

	To deliver a system that is consistent and of high quality, practitioners need to take a large number of quality characteristics into account in development~\cite{Chung2009a}. 
	Some of these characteristics are internal or relate to the development process and mainly concern developers (e.g., traceability) while others such as performance and usability are critical for end users~\cite{ISO250102011}. 
	Usability  is defined as \textit{``the extent to which a system, product or service can be used by specified users to achieve specified goals with effectiveness, efficiency and satisfaction in a specified context of use.''}~\cite{ISO9241}.
	At a more abstract level,  the actual experience of the end users with a piece of software also needs to be taken into account.
	This has led to introducing and studying the concept of User eXperience (UX): \emph{a user's holistic experience and perception of functionalities and quality characteristics of a piece of software}~\cite{Hassenzahl2003a}. 
	Researchers emphasize that developers cannot necessarily create the intended experience for the end users (e.g. feeling scared in a video game, or motivated in an e-learning system) merely thorough assuring usability~\cite{Hassenzahl2010a}.

	Nevertheless, studies show that software companies often face various challenges in their work with UX.
	Among other things, researchers relate these challenges to practitioners' low knowledge and awareness of UX and low industrial impact of UX theories~\cite{Lallemand2014a,Kashfi2016a}. 
	This can be addressed at least partially by developing suitable practical UX models~\cite{Chung2009a,Kashfi2016a}.
	Models can be formal (e.g., analytical) or informal (e.g.,  conceptual).
	In this study, we developed a conceptual requirement model that presents the interrelation between UX, functional and quality requirements.	
	
	We focused on requirements because they play an important role in effective practice of UX.
	For instance, Ardito et al.~\cite{Ardito2014a} empirically show that if practitioners fail to include UX in requirements documents, UX practices often become neglected in projects~\cite{Ardito2014a}.
	Similarly, Lanzilotti et al.~\cite{Lanzilotti2015a} argue that if UX is excluded from requirements documents,  often limited or no resources get assigned to UX work.

	Our model mainly targets software development practitioners, especially those who have little or no UX background and experience.
	The main goal of the model is to
	(i) help increasing practitioners' knowledge and awareness of UX, and
	(ii)  facilitate overcoming current UX-related communication gap among practitioners.
	We aim to achieve these goals through providing a common terminology that is familiar to and understandable for practitioners with both Software Engineering (SE) and Human Computer Interaction (HCI) backgrounds.
	
	Admittedly, various UX models have been developed so far, mainly in the field of HCI~\cite{Hassenzahl2003a,Wright2010a}.
	But such models are often too complex and use terminologies less familiar to practitioners with SE or similar technical backgrounds~\cite{Kashfi2012}.
	In addition, these models do not clearly present the interrelation between UX and other software quality characteristics and their corresponding models (e.g., ISO/IEC~25010).
	For instance, Hassenzahl~\cite{Hassenzahl2003a} discusses how utility (i.e., relevant functionality) and usability contribute to achieving a better UX.
	However, his model lacks references to other quality characteristics and makes no explicit connection to other software quality models or standards.
	Through mapping UX models and concepts to models and standards in SE and using similar terminologies as them, we can facilitate a better understanding of UX among practitioners with more technical backgrounds.
	
	In the field of SE as well, there have  been efforts to model the concept of UX as an emerging software quality characteristic.
	Some researchers have focused on extending ISO/IEC  standards on software quality models to incorporate UX~\cite{Bevan2008a}.
	In  ISO/IEC~25010, concepts related to UX are included in the definition of \emph{Quality in Use} (QiU):
	\textit{``the degree to which a product or system can be used by specific users to meet their needs to achieve specific goals with effectiveness, efficiency, freedom from risk and satisfaction in specific contexts of use.''}
	Similar to UX, QiU also emphasizes users' personal (aka. non-task-related) needs and emotional reactions, and includes `pleasure' (i.e., an emotional consequence of interacting with a piece of software) as a quality characteristic.
	In ISO/IEC~25010, usability is a part of \emph{Product Quality} (PQ) model.
	This model includes properties of the software product and computer system that determine the quality of the product in particular contexts of use.
	According to this standard, PQ affects QiU, i.e., the experience of users.
		
	Both Hasssenzahl's model of UX~\cite{Hassenzahl2003a} and ISO/IEC~25010 software quality model~\cite{ISO250102011}  are well established in HCI and SE communities respectively.
	Therefore, our model is inspired by these two models.
	Our model presents a categorization of quality requirements based on whether they can be measured objectively or not.
	To the best of our knowledge, current requirements literature does not include such a categorization.
	Our model aims to be a descriptive, simple,  practical, and actionable model for practitioners rather than a contribution to UX models and theories.
	
	This paper presents our model and the results of its initial validation through interviews with researchers and practitioners.
	Section~\ref{sec:method} describes our methodology.
	Section~\ref{sec:results} presents the model and our analysis of the interview data.
	Section~\ref{sec:disc} includes the discussion and ends with our  conclusion and suggestions for future research.

	\section{Research Approach}
	\label{sec:method}
	Our model was developed in close collaboration with industry.
	We followed the steps suggested by Gorschek et. al.~\cite{Gorschek2006a} in their \emph{technology transfer model}:
	\begin{itemize}
		
	\item  \textbf{problem issue in industry}:
	as elaborated in Section~\ref{sec:intro}, we were motivated by previous empirical findings on challenges with UX work in software industry; and that many of these challenges relate to practitioners' lack of knowledge and awareness of UX.
	
	\item  \textbf{study state of the art and problem formulation}:
	the model was developed based on ample literature study on UX and software quality characteristics.
	Two main models that inspired our work are Hassenzahl's UX model~\cite{Hassenzahl2003a} and the most recent ISO/IEC  standard on software quality~\cite{ISO250102011}.
	
	\item  \textbf{candidate solution}:
	in a series of workshops, the authors developed and refined a UX-aware model of requirements.
	
	\item  \textbf{validation in academia}:
	validation in academia was performed through interviews with four researchers.
	Two of the researchers have a SE background and the other two a HCI background with focus on UX. 
	
	\item  \textbf{static validation in industry}: 
	for initial industrial (i.e., static) validation in industry, we interviewed eight practitioners with different backgrounds, from four companies. 
	
\end{itemize}
	
	We selected our industrial interviewees based on their backgrounds and roles in the companies. 
	Four of them represent technical roles (e.g., developers and management with technical background) and four represent design roles (e.g.,  interaction designers and management with design background).
	This served to validate the model from two different perspectives: SE and HCI.
	When quoting the interviewees,  we did not include their role titles since we did not see a noticeable difference among the views in relation to the roles.
	Instead, to emphasize the views in relation to the two communities that our model targets, the quotes are marked with either SE or UX.
		
	The interviews were performed individually, face-to-face, and lasted between 30 to 60 minutes. We chose semi-structured interviews~\cite{Runeson2008a} to collect more of the interviewees' viewpoints and reflections.
	For this purpose, an interview guide was developed that included five main questions about correctness and understandability of the model (e.g. are the definitions provided by the model clear? how do they relate to your understanding of these concepts?)
		
	 In our study, we also paid attention to validity threats~\cite{Runeson2008a}.
	To increase construct validity (i) we minimized selection bias by selecting the subjects based on their role and experience, and (ii) we minimized the influence of researcher's presence on the behavior and response of the subjects by guaranteeing  the confidentiality of the data.
	To increase internal validity, we recorded the interviews in audio format, and in three cases in form of extensive notes.
	To increase external validity,  we sampled a number of different organizations in different industrial domains.
	However, since the interviews are just a sample they should be interpreted with some caution.

\section{Results and Analysis}
\label{sec:results}
\begin{figure*}
  \centering
\includegraphics[angle=0,scale=0.46]{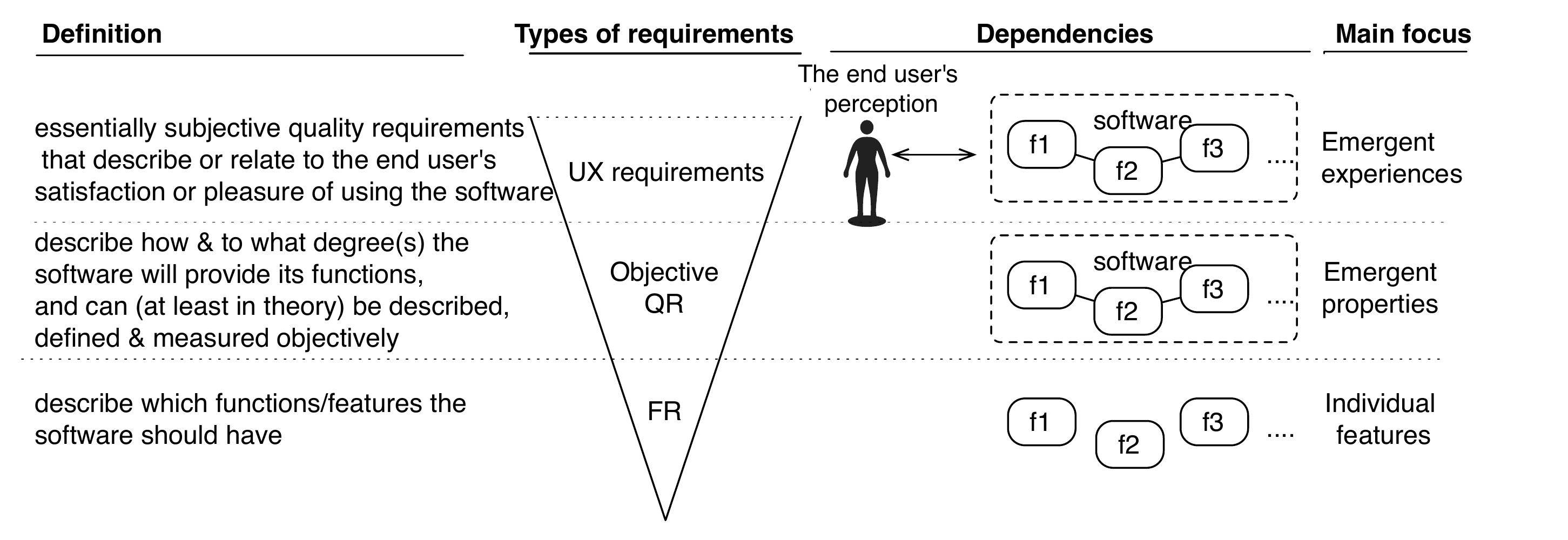}
\caption{A UX-aware model of requirements}
	\label{fig:reqs}
\end{figure*}

As figure~\ref{fig:reqs} depicts, our model introduces the concept of \textit{UX requirements} and puts it in relation to two other requirement types: \textit{objective Quality Requirements} (objective QRs) and Functional Requirements (FRs).
The model also includes definitions of these different requirement types.
UX requirements cover aspects such as usability, usefulness, emotions, aesthetics, motivations, and values.
For instance, `the end user shall feel in control' (emotions), `the system shall have a minimalistic design' (aesthetics), `the system shall facilitate getting quick access to trendy news' (motivations), `the system shall advocate recycling' (values).

Our model is presented using a reverse pyramid to emphasize that higher layers emerge from and depend on requirements below.
For example, an objective QR that describes performance needs to be stated in relation to some (or sets of) specific functions or features on which the performance is to be measured.
Thus, it assumes some FRs have already been (or at least could have been) established.
This is why QRs are often known to be \emph{cross-cutting}.
Similarly, a user's perception of the software (i.e., UX) can be constrained by UX requirements but implies some FRs or objective QRs that the perception is based on.
UX literature emphasizes this by highlighting the emergent nature of UX~\cite{Hassenzahl2003a}.
We stress that the use of layers does not mean one should first consider or implement the lower levels of requirements.
Also, the size of the areas do not reflect the quantity or significance of different requirement types.

In our model, we divide QRs into two categories of objective and subjective.
We emphasize that both FRs and objective QRs can be evaluated objectively  (i.e. measured/tested) without reference to a specific end user.
On the contrary, a group of requirements are subjective and should be singled out among the QRs.
Since these requirements always involve \emph{users' subjective perception}, we call them UX requirements.
We note that in practice, objective QRs often can also involve subjectivity since it is not cost-effective to specify them to a degree that they are fully objectively measurable.
This means that the subjectivity of these requirements is \emph{accidental}\footnote{The terms essential and accidental were originally used by Aristotle, and later adopted in the context of software development by Brooks~\cite{Brooks1987a} in his classification of complexities in software engineering.}.
On the other hand, UX heavily relies on human perception and is essentially subjective~\cite{Hassenzahl2003a} .

The role of \emph{human perception} (and therefore subjectivity) increases as we move upwards in the model.
For instance, a user may perceive particular features of software to be secure while another user may perceive the same features as insecure.
In addition, the level of \emph{abstraction} typically increases as we move upwards in the model. 
For instance `shall evoke a sense of trust' is a more abstract concept compared to `shall be secure' (objective QR) or `shall have a log-in function' (FR).

UX of a piece of software, among other aspects, emerges from underlying functionalities and objective quality characteristics (i.e., objective QRs), and the user's perception of them in each certain situation~\cite{Hassenzahl2003a}.
A designer can select a group of specific functionalities to increase the likelihood of creating a particular experience for the end users~\cite{Hassenzahl2003a}.
To emphasize the emergent nature of UX, we used a reverse pyramid in our model.
Putting UX requirements on top highlights that UX emerges from the underlying functionalities and quality characteristics.
For instance, in order to be trustworthy (abstract) the system provides a good overview of the functions available (concrete).
This resembles the cross-cutting nature of other quality characteristics.
Researchers emphasize that although practitioners may manipulate UX through these underlying elements, they still cannot guarantee a certain overall UX~\cite{Hassenzahl2003a,Wright2010a}.

The model was validated through interviews with eight software practitioners and four researchers.
All of the interviewees were positive regarding clarity and understandability of the model.
For instance, one of the interviewees said:  \textit{``My first impression of the model is that it is clear and easy to read.
	It is easy to understand what UX is and what extra `things' are needed to make more UX-aware decisions.''} (SE).
The participants had some suggestions regarding the terms and shapes used in the model.
These suggestions were taken into account when revising the model to the version we have presented above. 
From the interviewees' perspective, the main potential use and benefits of the model are as follows:

\textbf{Raising knowledge and awareness of various requirement types}
The interviewees stated that the model can raise knowledge and awareness of the role of all requirement types in achieving the intended UX.
Pointing to the two bottom layers of the model, one of the interviewees stated: \textit{``You can define something that looks really cool [\ldots] but to consistently deliver a good UX, we need to go the whole way down.''} (HCI).
Moreover, the interviewees generally agreed that to achieve the intended UX, FRs and objective QRs are important but not enough.
In their view, the model clearly presents this matter. 
In addition, according to the practitioners, the two communities still disagree on the importance of viewing quality characteristics from not only the system perspective but also the end users' perspective.
Regarding this a designer stated: \textit{``We have quite an argument with technical people because [in our view] the perceived performance is more important than the actual performance, usually.''} (HCI).
Some practitioners with SE background believed UX requirements can be treated the same as other types of requirements:
\textit{``The practical application of discussions, elicitation, specifying UX goals and UX requirements, all of this is something we already do for any other goals and requirements.''} (SE).
This contradicts the view of practitioners with UX knowledge: \textit{``[SE people] go through emotions and have it in their check lists, but it is not at the center of their effort [\ldots]. That's perfectly fine when you work with the functional level, but there are tons of other complexities that you need to consider.''} (HCI).

\textbf{Raising knowledge and awareness of UX-aware testing}
The concept of testing and its challenges was repeatedly brought up by the interviewees.
They generally agreed that quantitative methods are insufficient for UX evaluation. 
For instance, one reason is that while they can identify the problem areas in design, they cannot explain why these problems exist.
They, therefore, cannot sufficiently inform the re-design of the software.
Nevertheless, as the interviewees highlighted, the field of SE puts more emphasis on quantitative methods.
Regarding this, one interviewee stated: \textit{``I think we have a problem that we have not addressed yet. When we write our requirements specifications we think all requirements should be testable either by a unit test, product test or system test; and subjective requirements are very hard to test, so I think we tend not to include them in our requirements specifications.''} (SE).
 
\textbf{Facilitating UX-related communication}
The interviewees highlighted the model can improve communication among stakeholders through a common terminology that is understandable for stakeholders with both SE and HCI backgrounds.
One of the interviewees stated:
 \textit{``a common terminology among the staff will improve the communication, particularly between us and the managers.''} (HCI).
In addition, presenting the model to practitioners opened up a series of discussions about how the companies support different requirement types in their current practices. 

\section{Discussion} 
\label{sec:disc}
	Current software quality and UX models are evidently not practical or actionable~\cite{Kashfi2016a,Lallemand2014a,Folstad2010a}.
	Therefore, we saw a need for a  practical model that can summarize and clarify the current UX models and connect them to software quality models.
	Requirements play an important role in effective UX work~\cite{Ardito2014a,Lanzilotti2015a}; thus our model focuses on requirements.
	
	In order to overcome the shortcomings of the current UX models, our model clearly situates UX requirements in relation to FRs and other QRs.
	The model is also simple, clear, and understandable for practitioners with both SE and HCI background as our initial validation shows.
	
	The model also sheds light on UX-aware elicitation and documentation of requirements.
	By introducing the notion of \textit{UX requirements} the model explicitly groups those quality requirements that are essentially subjective and relate to the end users' perception.
	We performed an initial validation of the model through interviews with researchers and practitioners.
	The validation confirmed correctness of the model, and that it can facilitate enhancing   knowledge and awareness of UX and UX-related communication among practitioners.
	
	In contrast to the approach taken in ISO/IEC~25010, we separated UX requirements from other QRs in our model.
	The reason was to emphasize that UX requirements are \textit{essentially subjective}, and separate them from \textit{accidentally subjective} quality requirements, what we call objective QRs. 
	By doing so, the model can extend and complement the current models of UX and software quality. 
	We have summarized our view on subjectivity and objectivity of different requirement types in Table~\ref{tbl:obj_subj}.
	\begin{table}[h]
		\centering
		\begin{tabular}{l|c|c|c}
			&functional requirements  & UX requirements & \specialcell{objective quality \\ requirements} \\
			\hline
			essentially objective & Yes &  &  Yes \\
			\hline
			essentially subjective & & Yes & \\
			\hline
			accidentally subjective & & &  Yes\\
			\hline
			\specialcell{possible to \\ evaluate objectively}  & Yes & 
			&  Yes\\
			\hline
			\specialcell{objectively evaluated \\ in practice} & Yes &   
			&  sometimes\\
			\hline
			\specialcell{subjectively evaluated \\ in practice} &   & Yes 
			&  sometimes\\
		\end{tabular}
		\caption{Differences in subjectivity and objectivity of various requirement types, and how they are treated in practice}
		\label{tbl:obj_subj}
	\end{table}
	
	FRs are objective by nature: we can objectively evaluate  whether a piece of software satisfies a specific FR or not.
	This is a binary evaluation: either a functionality is implemented in the software or not.
	On the contrary, QRs (including usability) are known to be more difficult to evaluate.
	This has led practitioners to often evaluate QRs subjectively and based on their personal judgment~\cite{Chung2009a}.
	Still, this does not mean that these requirements are not possible to be evaluated objectively.
	Therefore, in our model we call them `objective QRs' and emphasize that they are essentially objective but still in practice accidentally subjective.
	
	If a requirement is subjective by accident, this means that the subjectivity is not a result of its nature but other reasons such as lack of knowledge and awareness, tools and methods, or costs.	
	In theory, it is possible and even recommended to evaluate these requirements objectively.
	Accidental subjectivity can be overcome as the field of requirements engineering matures.
	For instance, by developing more tools and methods to facilitate measuring  these requirements objectively (e.g.~\cite{Gilb2005a,SvenssonBerntsson2015a}).

	\begin{sidewaystable}[!htbp]
		\centering
		\footnotesize
		\begin{tabular}{|p{0.22\textwidth}|p{0.39\textwidth}|p{0.39\textwidth}|}
			\hline
			\textbf{Characteristics of UX} &
			\textbf{Implications for practice} &
			\textbf{Open research problems} \\ \hline
			\emph{Abstract and emergent}:
			
			experience emerges from underlying 
			functionalities and objective quality characteristics &
			\begin{itemize}[topsep=0pt,leftmargin=*]
				\item practitioner need to identify UX requirements and refine them into concrete FRs and 
				objective QRs 
				
				(sample method in \cite{Hassenzahl2001a})
				
				\item practitioners need to evaluate UX both holistically and via  evaluating its underling elements, i.e., users' perception of objective QRs and FRs 		
				
				(for a summary of evaluation methods see~\cite{Zimmermann2008a})	
			\end{itemize} 
			&
			\begin{itemize}[topsep=0pt,leftmargin=*]
				\item UX requirements are difficult to refine and translate into design solutions and more concrete requirements; there are limited guidelines to support that
				
				\item 	still, there is no standardized and agreed upon set of UX 		measures and metrics
				
				\item 	there are limited guidelines on how to choose suitable UX measures and metrics and interpret their findings to improve the overall UX

			\end{itemize}\\ \hline
			\emph{Essentially subjective}:
			experience heavily relies on human perception therefore  is essentially  subjective &
			\begin{itemize}[topsep=0pt,leftmargin=*]
				\item 	qualitative user opinion should be used in evaluations  
				
				(for a summary of evaluation methods see~\cite{Zimmermann2008a})	
				
				\item  when measuring UX, practitioners need to involve  statistically significant number of users to guarantee reliable data
				 (for more information see~\cite{Zimmermann2008a,Law2014b})

			\end{itemize}
			&
			\begin{itemize}[topsep=0pt,leftmargin=*]
				\item there are limited theories  on the relationship between UX and memory 
				
			\end{itemize} \\ \hline
			\emph{Temporal}:
			
			experience can change over time
			& 
			practitioners need to assure that the relation between time and experience is reflected in requirements testing activities & 
			\begin{itemize}[topsep=0pt,leftmargin=*]
				\item  current body of knowledge includes limited theories on the relationship between UX and time
				\item  practitioners have limited access to tools and methods to handle temporality in evaluation
			\end{itemize} 
			\\ \hline
		\end{tabular}
		\caption{Characteristics of UX and their implication for practice of software development and future research, especially concerning UX requirements }
		\label{tbl:uxr}
	\end{sidewaystable}

	In contrast to FRs and objective QRs, UX requirements are essentially subjective.
	UX heavily relies on human perception and is therefore by nature subjective.
	Even in cases when UX is measured, the measurement is an approximation of the real experience of users.
	Especially since the phenomenon of experience is prone to fabrication and fading since it heavily relies on human memory~\cite{Law2014b}.
	Still, practitioners can approximately measure UX through gathering users' opinions, for instance using questionnaires (e.g. AttrakDiff, Self-assessment Manikin, the affect gird~\cite{Zimmermann2008a}).
	For an overview of various approaches to UX evaluation and measurement, we can refer to Law et al.~\cite{Law2014b} and Zimmermann~\cite{Zimmermann2008a}.
	 
	When measuring UX, statistically significant number of heterogeneous users need to be involved to guarantee reliable results~\cite{Law2014b}.
	In contrast to UX requirements, practitioners can test objective QRs even 
		without involving users (e.g., automatically).
		For instance, practitioners can automatically compute usability measures by running a user interface specification through some program~\cite{Nielsen1994a}.
	UX requirements also differ from objective QRs in that their metrics and measures are not agreed upon or standardized yet; that makes their measurement even more difficult.
	On the other hand, for objective QRs (including usability) practitioners have access to relevant standards, e.g., ISO/IEC~9126~\cite{ISO9126-4-2004}.
	
	The \emph{emergent nature of UX} can partially explain why practitioners and researchers still do not agree on UX metrics and measures.
	For example, Law et al.~\cite{Law2014b} empirically show that often practitioners and researchers have two different attitudes towards UX measurement.
	They are either strongly convinced that it is ``necessary, plausible and feasible'' to measure UX through its finest underlying elements, or doubtful about the ``necessity and utility''  of measuring these elements.
	Law et al. further discuss that practitioners do not still have enough guidelines on how to choose suitable UX measures and metrics to measure these elements or to interpret the findings to better re-design the software~\cite{Law2014b}.
	
	We identified at least one more issue that relates to the abstract and emergent nature of UX: 
	practitioners still do not have enough support for refining UX requirements to more concrete design solutions and requirements (i.e. FRs and objective QRs)~\cite{Kashfi2016a}.
	One of the few existing methods for a UX-aware requirements work is developed by Hassenzahl~\cite{Hassenzahl2001a}.
	Hassenzahl~\cite{Hassenzahl2001a} emphasizes that, in their work with UX, practitioners should refine the abstract requirements into functionalities and concrete quality characteristics.
	He further emphasizes that this should be performed in close collaboration with the end users' representatives.

	\emph{Temporality} is another important characteristic of UX that differentiates UX requirements from objective QRs.
	Temporality implies that experience of a user with a piece of software can change over time~\cite{Hassenzahl2003a}.
	Researchers therefore recommend practitioners to take the whole spectrum of interaction into account when designing or evaluating the UX of a piece of software~\cite{Wright2005}.
	Practitioners should pay attention to the users' experiences not only during, but also before and after the interaction~\cite{Hassenzahl2003a,Wright2010a}.
	Thus, UX requirements should also reflect the spectrum of experience.
	For instance, a UX requirement may concern users' first impression: \textit{``average score of responses to questionnaire questions on initial impression and satisfaction should be higher than X.''}
	Another requirement may concern users' overall experience: \textit{``average score of responses to questionnaire questions should be higher than X.''}
	In contrary, FRs and objective QRs are not dependent on time.
	For instance, practitioners get the same results if they repeat measuring performance or security of the software over time (providing that the software and the test context, e.g., CPU load, have not changed).
	Table~\ref{tbl:uxr} summarizes the main characteristics of UX and how they lead to differences between UX and other requirement types.

	As a key initial step, to perform UX-aware requirements and evaluation work, practitioners require to understand the differences between UX requirements and objective QRs.
	But there are still a number of important issues that need attention to and plan for improvements.
	For instance, to facilitate a \emph{UX-aware requirement elicitation}, practitioners require knowledge and awareness of human psychological needs and their relation to `experiences'.
	They need to know what to look for and how to look for it.
	To facilitate a \emph{UX-aware requirements documentation}, practitioners need to have access to tools, methods and guidelines on how to document and communicate the results of elicitation in form of various UX requirements.
	In addition, these tools and methods should be integrated into current requirements tools and methods.
	To facilitate a \emph{UX-aware verification and validation}, practitioners need to have access to suitable tools, methods and guidelines that can help investigating whether these requirements are satisficed or not.
	Other open research problems concern traceability, conflict resolution, prioritization, and cost-estimation of UX requirements.
	
	To start investigating how to better support UX requirements in practice, we suggest the communities to first investigate current tools, methods and guidelines for supporting  usability in the above activities.
	We do not however claim that current tools, methods and guidelines for supporting usability are established and flawless; for the purpose we suggest here they do not need to be so.
	Since UX and usability are related, we believe we can get inspired by and learn from usability literature since it is comparatively more mature.
	Still, we need to pay attention to essential differences between the two concepts.

	We hope to have convinced the reader that UX in general, and UX requirements in particular are worth pursuing in software development research and practice.
	We facilitate this through explicitly separating essentially subjective UX requirements from other requirement types, and raising knowledge and awareness of these requirements.
	However, as we mentioned, there are still a number of open research questions that the communities need to address.
	We also hope to have inspired extending current software quality and requirements models and standards to better support the concept of UX.	
	Future research should introduce the model to software development companies, provide even more detailed advice and examples on how to elicit, document, and break down UX requirements and refine them to other more concrete requirement types.

\bibliographystyle{splncs}
\bibliography{library.bib} 
\end{document}